\documentstyle[12pt]{article}
\begin{document}

\title{Some analytical models of radiating collapsing spheres}

\author{ L. Herrera$^1$\thanks{e-mail:
laherrera@movistar.net.ve},A Di Prisco$^1$\thanks{e-mail:
adiprisc@fisica.ciens.ucv.ve} and J. Ospino$^{2}$\thanks {e-mail:
jhozcrae@usal.es} \\
\small{$^1$Escuela de F\'{\i}sica, Facultad de Ciencias,} \\
\small{Universidad Central de Venezuela, Caracas, Venezuela.}\\
\small{$^2$Area de F\'\i
sica Te\'orica. Facultad de Ciencias,Universidad de Salamanca}\\
\small{Salamanca, Spain.}\\
}

\maketitle

\begin{abstract}
We present some analytical solutions to the Einstein equations, describing radiating collapsing spheres in the diffusion approximation. Solutions allow for modeling physical reasonable
situations. The temperature is calculated for each solution, using a hyperbolic transport equation, which permits to exhibit the influence of relaxational effects on the dynamics
of the system.
\end{abstract}

\maketitle

\newpage

\section{Introduction}
Our purpose in this work is to provide some analytical solutions to Einstein equations, describing
collapsing dissipative  spheres in the difussion approximation. Such solutions may serve as test--bed for numerical relativity, and for probing cosmic censorship and hoop
conjecture, among other important issues, and
represent a natural tool to  bring out the influence of dissipation on the evolution of a collapsing object. 

Analytical solutions although  generally are
found, either for too simplistic equations of state and/or under additional
heuristic assumptions whose justification is usually uncertain, are more suitable for a
general discussion than  purely
numerical solutions, which usually hinder to catch qualitative, aspects
of the process.

Therefore it seems useful to consider analytical models which are relatively
simple to analyze but still contain some of the essential features of a
realistic situation.

Our endeavour is further justified by the following two considerations:
\begin{itemize}
\item It is already an established fact, that gravitational collapse is a
highly dissipative process (see \cite{Hs}, \cite{Hetal}, \cite{Mitra} and references
therein). This dissipation is required to account for the very large
(negative) binding energy of the resulting compact object (of the order
of $-10^{53} erg$)

 Indeed, it appears that the only plausible
mechanism to carry away the bulk of the binding energy of the collapsing
star, leading to a neutron star or black hole is neutrino emission
\cite{1}.

\item In the diffusion approximation, it is assumed that the energy flux of
radiation (as that of
thermal conduction) is proportional to the gradient of temperature. This
assumption is in general very sensible, since the mean free path of
particles responsible for the propagation of energy in stellar
interiors is in general very small as compared with the typical
length of the object.
Thus, for a main sequence star as the sun, the mean free path of
photons at the centre, is of the order of $2\, cm$. Also, the
mean free path of trapped neutrinos in compact cores of densities
about $10^{12} \, g.cm.^{-3}$ becomes smaller than the size of the stellar
core \cite{3,4}.

Furthermore, the observational data collected from supernovae 1987A
indicates that the regime of radiation transport prevailing during the
emission process, is closer to the diffusion approximation than to the
streaming out limit \cite{5}.

Accordingly we shall restrict here to this later case, being aware that
there are situations in stellar evolution where that approximation fails.
\end{itemize}

During their evolution, self--gravitating objects may pass
through phases of intense dynamical activity, with time scales of the order
of magnitude of (or even smaller than) the hydrostatic time scale, and  for
which the quasi--static
approximation is clearly not reliable, e.g.,the collapse of very massive
stars \cite{8''}, and the quick collapse phase
preceding neutron star formation, see for example \cite{9'} and
references therein. In these cases it is mandatory to take into account
terms which describe departure from
equilibrium, i.e. a full dynamic description has to be used.

Here we are mainly concerned with the quick collapse phase, which  
implies that we have to  appeal to a hyperbolic theory of
dissipation.	The use of a hyperbolic theory of dissipation is further justified  by the
necessity of overcoming the difficulties inherent to parabolic theories
(see references \cite{Hs}, \cite{18}--\cite{8'} and references therein). Doing so we shall be able to give a description of processes occuring before thermal relaxation.

 Many analytical solutions of Einstein's field
equations with dissipative fluids carrying heat flow have been studied (see
\cite{Bonnor} for references
up to
1989 and \cite{relax}, \cite{GM} for more recent ones).

In this vein here we present some models of conformally flat dissipative spherical collapse with shear-free
motion.  We match our models  to a radiating null field described by the outgoing Vaidya spacetime.
 
Although the shear free and the conformally flat conditions are introduced
here in order to simplify calculations, it is worth noticing that these
conditions generalize physical assumptions widely used in astrophysics.
Indeed, the shear free condition in the Newtonian regime describes the
homologous evolution and the role of shear has been extensively considered in general
relativity \cite{SFCF}. On the other hand it is well known that the
conformally flat condition implies in the perfect fluid
case the homogeneity of the energy density distribution \cite{Hetal}.

The paper is organized as follows. In section 2 the field equations, conventions and junction conditions are
presented; in section 3 we present the general solution corresponding to the 
conformal flatness condition; in sections 4 and 5, particular analytical solutions are given, and  a specific model is constructed from one of them;
finally a brief conclusion is presented.

\section{The fluid distribution and the interior spacetime}
We assume a sphere of collapsing perfect fluid with heat flow. Its spherical
surface $\Sigma$ has center 0 and is filled with radially moving perfect
fluid conducting heat flow, so having energy momentum tensor
\begin{equation}
T_{\alpha\beta}=\left(\mu+p\right)w_{\alpha}w_{\beta}+pg_{\alpha\beta}+
q_{\alpha}w_{\beta}+w_{\alpha}q_{\beta}, \label{I1}
\end{equation}
where $\mu$ and $p$ are the proper density and pressure of the fluid,
$w_{\alpha}$ its unit four-velocity, $q_{\alpha}$ the heat conduction
satisfying $q_{\alpha}w^{\alpha}=0$ and $g_{\alpha\beta}$ is the metric
tensor of spacetime.

We choose comoving coordinates within $\Sigma$ and impose shear-free fluid
motion
which allows the metric be written in the form (see \cite{Herrera} for details)
\begin{equation}
ds^2=-A^2dt^2+B^2\left[dr^2+r^2\left(d\theta^2+\sin^2\theta
d\phi^2\right)\right],
\label{I2}
\end{equation}
where $A$ and $B$ are only functions of $r$ and $t$. We number the
coordinates $x^0=t$, $x^1=r$, $x^2=\theta$ and $x^3=\phi$ and then we
have the four-velocity given by
\begin{equation}
w_{\alpha}=-A\delta_{\alpha}^0, \label{I3}
\end{equation}
and the heat flows radially,
\begin{equation}
q^{\alpha}=q\delta_1^{\alpha}, \label{I4}
\end{equation}
where $q$ is a function of $r$ and $t$. In these coordinates the equation of the boundary surface $\Sigma$ is given by $r=r_{\Sigma}=constant$.

The spacetime described by (\ref{I2}) has the following non-null components
of the Weyl tensor $C_{\alpha\beta\gamma\delta}$,
\begin{equation}
C_{2323}=\frac{r^4}{3}B^2\sin^2\theta\left[\left(\frac{A^{\prime}}{A}-
\frac{B^{\prime}}{B}\right)\left(\frac{1}{r}+2\frac{B^{\prime}}{B}\right)-
\left(\frac{A^{\prime\prime}}{A}-\frac{B^{\prime\prime}}{B}\right)\right],
\label{I6}
\end{equation}
and
\begin{eqnarray}
C_{2323}=-r^4\left(\frac{B}{A}\right)^2\sin^2\theta C_{0101}
=2r^2\left(\frac{B}{A}\right)^2\sin^2\theta C_{0202} \nonumber \\
=2r^2\left(\frac{B}{A}\right)^2C_{0303}=-2r^2\sin^2\theta C_{1212}=-2r^2C_{1313},
\label{I6a}
\end{eqnarray}
where the primes stand for differentiation with respect to $r$ and dot stands for differentiation with respect to $t$.
\subsection{The field equations}
The non null components of Einstein's field equations
$G_{\alpha\beta}=8\pi T_{\alpha\beta}$, where $G_{\alpha\beta}$ is the
Einstein tensor and $T_{\alpha\beta}$ is given by (\ref{I1}), with metric
(\ref{I2}) are
\begin{eqnarray}
G_{00}=-\frac{A^2}{B^2}\left[2\frac{B^{\prime\prime}}{B}-
\left(\frac{B^{\prime}}{B}\right)^2+\frac{4}{r}\frac{B^{\prime}}{B}\right]+
3\left(\frac{\dot{B}}{B}\right)^2=8\pi \mu A^2, \label{I7} \\
G_{11}=\left(\frac{B^{\prime}}{B}\right)^2+\frac{2}{r}\frac{B^{\prime}}{B}+
2\frac{A^{\prime}}{A}\frac{B^{\prime}}{B}+\frac{2}{r}\frac{A^{\prime}}{A}
\nonumber \\
+\frac{B^2}{A^2}\left[-2\frac{\ddot{B}}{B}-\left(\frac{\dot{B}}{B}\right)^2+
2\frac{\dot{A}}{A}\frac{\dot{B}}{B}\right]=8\pi p B^2, \label{I8} \\
G_{22}=\frac{G_{33}}{\sin^2\theta}=r^2\left[\frac{A^{\prime\prime}}{A}+
\frac{1}{r}\frac{A^{\prime}}{A}+\frac{B^{\prime\prime}}{B}-
\left(\frac{B^{\prime}}{B}\right)^2+\frac{1}{r}\frac{B^{\prime}}{B}\right]
\nonumber \\
+r^2\frac{B^2}{A^2}\left[-2\frac{\ddot{B}}{B}-
\left(\frac{\dot{B}}{B}\right)^2
+2\frac{\dot{A}}{A}\frac{\dot{B}}{B}\right]=8\pi p r^2 B^2, \label{I9} \\
G_{01}=-2\left(\frac{\dot{B}}{AB}\right)^{\prime}A=-8\pi q A B^2. \label{I10}
\end{eqnarray}

The mass function $m(r,t)$ of Cahill and McVittie \cite{Cahill} is obtained
from the Riemann tensor component ${R_{23}}^{23}$ and it is
for metric (\ref{I2})
\begin{equation}
m(r,t)=\frac{\left(rB\right)^3}{2}{R_{23}}^{23}=\frac{r^3B}{2}\left[
\left(\frac{\dot{B}}{A}\right)^2-
\left(\frac{B^{\prime}}{B}\right)^2\right]-r^2B^{\prime}. \label{I11}
\end{equation}

For studying the dynamical properties of the field equations and following Misner and Sharp \cite{MS}, let us  introduce  the velocity $U$ of the collapsing fluid as
\begin{equation}
U=rD_tB(<0 \;\; in \; the \;  case \; of \; collapse), \label{19N}
\end{equation}
where
the proper time derivative $D_t$, is given by
\begin{equation}
D_t=\frac{1}{A}\frac{\partial}{\partial t}. \label{16N} 
\end{equation}
\subsection{Junction conditions}
If the collapsing fluid lies within a spherical surface $\Sigma$ it must be
matched to a suitable exterior. Since heat will be leaving the fluid across
$\Sigma$, the exterior is not vacuum, but the outgoing Vaidya spacetime which
models the radiation and has metric
\begin{equation}
ds^2=-\left[1-\frac{2m(v)}{\rho}\right]dv^2-2dvd\rho+\rho^2(d\theta^2+\sin^2
\theta
d\phi^2),
\label{III1}
\end{equation}
where $m(v)$ is the total mass inside $\Sigma$ and is a function of the retarded
time $v$. In (\ref{III1}) $\rho$ is a radial coordinate given in a
non-comoving frame.

The conditions for the matching of these two spacetimes (\ref{I2}) and (\ref{III1}), are the Darmois conditions \cite{Darmois}, which using the
field
equations (\ref{I8}-\ref{I10}) and the mass function (\ref{I11}) imply
\cite{Bonnor}
\begin{eqnarray}
p_\Sigma=(qB)_\Sigma, \label{Nu10} \\
(qB)_{\Sigma}=\frac{1}{4\pi}\left(\frac{L}{\rho^2}\right)_{\Sigma}, \label{Nu11} \\
(rB)_{\Sigma}=\rho_{\Sigma}, \label{12} \\
\left(\frac{r^3}{2}\frac{B\dot{B}^2}{A^2}-\frac{r^3}{2}\frac{B^{\prime 2}}{B}-r^2B^{\prime}\right)_{\Sigma}=m(v),
\label{13} \\
A_{\Sigma}dt=\left(1-\frac{2m}{\rho}+2\frac{d\rho}{dv}\right)^{1/2}_{\Sigma} dv, \label{Nu14}
\end{eqnarray}
where $L$ is defined as the total luminosity of the collapsing sphere as measured on its surface and is given by
\begin{equation}
L=L_{\infty}\left(1-\frac{2m}{\rho}+2\frac{d\rho}{dv}\right)^{-1}, \label{Nu14a}
\end{equation}
and where
\begin{equation}
L_{\infty}=\frac{dm}{dv} \label{Nu14b}
\end{equation}
is the total luminosity measured by an observer at rest at infinity.
\section{Conformally flat solution}
Here we impose conformal flatness to the spacetime given by (\ref{I2}),
i.e. all its Weyl tensor components must be zero valued. Then it can be shown that metric functions $A$ and $B$ take the form (see \cite{Herrera} for details)
\begin{equation}
A=\left[C_1\left(t\right)r^2+1\right]B, \label{II2bis}
\end{equation}
where $C_1$ is an arbitrary function of $t$
and
\begin{equation}
B=\frac{1}{C_2(t)r^2+C_3(t)}, \label{II4bis}
\end{equation}
where $C_2$ and $C_3$ are arbitrary functions of $t$.

Substituting solution (\ref{II2bis}) and (\ref{II4bis}) into (\ref{I8}),
(\ref{I9}) and (\ref{I11}) we obtain,
\begin{eqnarray}
8\pi\mu=3\left(\frac{\dot{C}_2r^2+\dot{C}_3}{C_1r^2+1}\right)^2+12C_2C_3,
\label{IIN6} \\
8\pi p=\frac{1}{(C_1r^2+1)^2}\left[2(\ddot{C}_2r^2+\ddot{C}_3)(C_2r^2+C_3)-
3(\dot{C}_2r^2+\dot{C}_3)^2 \right. \nonumber \\
\left. -2\frac{\dot{C}_1}{C_1r^2+1}
(\dot{C}_2r^2+\dot{C}_3)(C_2r^2+C_3)r^2\right] \nonumber \\
+\frac{4}{C_1r^2+1}\left[C_2(C_2-2C_1C_3)r^2+C_3(C_1C_3-2C_2)\right],
\label{IIN7}\\
8\pi q=4(\dot{C}_3C_1-\dot{C}_2)
\left(\frac{C_2r^2+C_3}{C_1r^2+1}\right)^2r. \label{IIN8}
\end{eqnarray}

Finally, from (\ref{Nu10}), (\ref{IIN7}) and (\ref{IIN8})   we have
\begin{eqnarray}
\left\{\ddot{C}_2r^2+\ddot{C}_3-\frac{3}{2}\frac{(\dot{C}_2r^2+\dot{C}_3)^2}
{C_2r^2+C_3}-
\frac{\dot{C}_1r^2(\dot{C}_2r^2+\dot{C}_3)}{C_1r^2+1}
-2(\dot{C}_3C_1-\dot{C}_2)r\right. \nonumber \\
\left. +2\frac{(C_1r^2+1)}{C_2r^2+C_3}
\left[C_2(C_2-2C_1C_3)r^2+C_3(C_1C_3-2C_2)\right]\right\}_{\Sigma}=0.
\label{III5}
\end{eqnarray}
Please note that a misprint in (\ref{III5}) appearing in \cite{Herrera} has been corrected here.

In the following sections we shall obtain some analytical solutions satisfying (\ref{Nu10})--(\ref{IIN8}).
\section{Solution I}
In order to integrate (\ref{III5}) let us assume
\begin{equation}
\left. \begin{array}{l}

C_2=\alpha C_3\\
C_1\equiv Const \\
\alpha=constant
\end{array} \right \}
\label{38}
\end{equation}
then replacing (\ref{38}) into (\ref{III5}) we get:
\begin{equation}
C_3\ddot C_3-\frac{3}{2}\dot C_3^2-\frac{2(C_1-\alpha)r_{\Sigma}}{\alpha
r_{\Sigma}^2+1} C_3\dot C_3+ \frac{2(C_1r_{\Sigma}^2+1)}{(\alpha
r_{\Sigma}^2+1)^2}[\alpha(\alpha-2C_1)r_{\Sigma}^2+(C_1-2\alpha)]C_3^2=0
\end{equation}
which, in terms of the new variable  $C_3(t)=u^{-2}(t)$, becomes:
\begin{equation}
\ddot u-\frac{2(C_1-\alpha)r_{\Sigma}}{\alpha r_{\Sigma}^2+1} \dot
u-\frac{(C_1r_{\Sigma}^2+1)}{(\alpha
r_{\Sigma}^2+1)^2}[\alpha(\alpha-2C_1)r_{\Sigma}^2+(C_1-2\alpha)]u=0
\end{equation}

The equation above allows the following three solutions:
\\
\\
\textbf{{Case I}} \qquad $(C_1-\alpha)^2 r_{\Sigma}^2+(C_1r_{\Sigma}^2+1)[\alpha
(\alpha-2C_1)r_{\Sigma}^2+(C_1-2\alpha)]>0$

\begin{equation}
\left. \begin{array}{l} C_3(t)=[\beta_1
e^{(\frac{(C_1-\alpha)r_{\Sigma}+\sqrt{(C_1-\alpha)^2r_{\Sigma}^2+(C_1r_{\Sigma}^2+1)[\alpha
(\alpha-2C_1)r_{\Sigma}^2+(C_1-2\alpha)]}}{\alpha r_{\Sigma}^2+1})t}
\\
\qquad \quad
+\beta_2e^{(\frac{(C_1-\alpha)r_{\Sigma}-\sqrt{(C_1-\alpha)^2r_{\Sigma}^2+(C_1r_{\Sigma}^2+1)[\alpha
(\alpha-2C_1)r_{\Sigma}^2+(C_1-2\alpha)]}}{\alpha r_{\Sigma}^2+1})t}]^{-2}
\end{array} \right \}
\label{C31}
\end{equation}
\\

\textbf{Case II }\qquad $(C_1-\alpha)^2r_{\Sigma}^2+(C_1r_{\Sigma}^2+1)[\alpha
(\alpha-2C_1)r_{\Sigma}^2+(C_1-2\alpha)]<0$
\begin{equation}
\left. \begin{array}{l}

C_3(t)=[e^{\frac{(C_1-\alpha)r_{\Sigma}}{\alpha r_{\Sigma}^2+1}t}(\beta_1
\cos(\frac{\sqrt{(C_1-\alpha)^2r_{\Sigma}^2+(C_1r_{\Sigma}^2+1)[\alpha
(\alpha-2C_1)r_{\Sigma}^2+(C_1-2\alpha)]}}{\alpha r_{\Sigma}^2+1})t
\\
\\
\qquad \quad
+\beta_2\sin(\frac{\sqrt{(C_1-\alpha)^2r_{\Sigma}^2+(C_1r_{\Sigma}^2+1)[\alpha
(\alpha-2C_1)r_{\Sigma}^2+(C_1-2\alpha)]}}{\alpha r_{\Sigma}^2+1})t)]^{-2}
\end{array} \right \}
\label{C32}
\end{equation}
\\
\textbf{Case III } \qquad $(C_1-\alpha)^2r_{\Sigma}^2+(C_1r_{\Sigma}^2+1)[\alpha
(\alpha-2C_1)r_{\Sigma}^2+(C_1-2\alpha)]=0$
\begin{equation}
C_3(t)=(\beta_1+\beta_2 t)^{-2}e^{-\frac{2(C_1-\alpha)r_{\Sigma}}{\alpha
r_{\Sigma}^2+1}t}
\label{C33}
\end{equation}

This solution reduces to the one found in \cite{GM} when
$\alpha =0$.
\subsection{Calculation of the temperature}
It is worth calculating the temperature  distribution, $T(r,t)$,
for our model,
through the Maxwell-Cattaneo heat transport equation. For simplicity we shall consider here the so called ``truncated'' version, for which it reads \cite{19}, \cite{8},
\begin{equation}
\tau
h^{\alpha\beta}w^{\gamma}q_{\beta;\gamma}+q^{\alpha}=-\kappa h^{\alpha\beta}
(T_{,\beta}+Ta_{\beta}), \label{V1}
\end{equation}
where $\tau$ is the relaxation time, $\kappa$ the thermal conductivity and
$h^{\alpha\beta}=g^{\alpha\beta}+w^{\alpha}w^{\beta}$ the projector orthogonal
to $w^{\alpha}$.
Considering (\ref{I2}--\ref{I4}) then (\ref{V1}) becomes
\begin{equation}
\tau(qB)\dot{}B+qAB^2=-\kappa(TA)^{\prime}. \label{V2}
\end{equation}

In our case the integration of (\ref{V2}) gives
\begin{equation}
\left. \begin{array}{l}
T(t,r)=[f(t)+\frac{\tau}{4\pi \kappa}\frac{\alpha
r^2+1}{C_1r^2+1}(\ddot C_3C_3+\dot C_3^2)
\\
\\
\qquad \qquad -\frac{\dot C_3}{4\pi
\kappa}ln(\frac{C_1r^2+1}{\alpha r^2+1})]\frac{\alpha r^2+1}{C_1
r^2+1}
\end{array} \right \}
\label{TN1}
\end{equation}

The integration function $f(t)$ may be easily related to the central temperature $T_c(t)$
\begin{equation}
f(t)=T_c(t)-\frac{\tau}{4\pi \kappa}(\ddot C_3C_3+\dot C_3^2)
\label{tempN}
\end{equation}
then we may write for the temperature

\begin{eqnarray}
\left. \begin{array}{l}
T(t,r)=[T_c(t)+\frac{\tau}{4\pi \kappa}\frac{(\alpha-C_1)
r^2}{C_1r^2+1}(\ddot C_3C_3+\dot C_3^2)
\\
\\
\qquad \qquad -\frac{\dot C_3}{4\pi
\kappa}ln(\frac{C_1r^2+1}{\alpha r^2+1})]\frac{\alpha r^2+1}{C_1
r^2+1}
\end{array} \right \}
\label{2TN}
\end{eqnarray}
where $C_3$ is given by either (\ref{C31}), (\ref{C32}) or (\ref{C33}).

In case $C_1=\alpha$ our system becomes a collapsing (non--dissipative) Friedmann dust sphere, as it can be checked from (\ref{IIN6})--(\ref{IIN8}). In this latter case the temperature,
as expected,  is homogeneous ($T(t)=T_c(t)$). Thus, models (\ref{C31})--(\ref{C33}) provide examples where inhomogeneity is directly related to dissipation.

\section{Solution II}
Another solution, with an interesting physical interpretation may be found by introducing the variable $v$, defined as
\begin{equation}
C_2(t)r_{\Sigma}^2+C_3(t)=v^{-2}(t)
\label{II1}
\end{equation}
into equation (\ref{III5}), obtaining:
\begin{equation}
\left. \begin{array}{l}
 -2v^{-2}(\frac{\ddot v}{v}-\frac{\dot C_1
r_{\Sigma}^2}{C_1r_{\Sigma}^2+1}\frac{\dot v}{v}+\frac{2\dot v}{r_{\Sigma}v})-\frac{2}{r_{\Sigma}}\dot
C_3 (C_1r_{\Sigma}^2+1)
\\
+\frac{2v^{2}(C_1r_{\Sigma}^2+1)}{r_{\Sigma}^2
}(v^{-4}-2v^{-2}C_3(2+C_1r_{\Sigma}^2)+3C_3^2(1+C_1r_{\Sigma}^2))=0
\end{array} \right \}
\label{II2}
\end{equation}
Then, defining
\begin{equation}
Z(t)=C_1r_{\Sigma}^2+1
\label{II3}
\end{equation}
equation (\ref{II2}) becomes:
\begin{equation}
\left. \begin{array}{l}
 -2v^{-2}(\frac{\ddot v}{v}-\frac{\dot Z
}{Z}\frac{\dot v}{v}+\frac{2\dot v}{r_{\Sigma}v})-\frac{2}{r_{\Sigma}}\dot C_3 Z
\\
+\frac{2Zv^{2}}{r_{\Sigma}^2}(v^{-4}-2v^{-2}C_3(1+Z)+3ZC_3^2)=0
\end{array} \right \}
\label{II4}
\end{equation}

Next, introducing the new variable $C(t)$ through $C_3(t)=v^{-2}C(t)$,  (\ref{II4}) may be written as:
\begin{equation}
\left. \begin{array}{l} \frac{\ddot v}{u}-\frac{\dot
Z}{Z}\frac{\dot v}{v}+\frac{2}{r_{\Sigma}}\frac{\dot
v}{v}-\frac{2Z}{r_{\Sigma}}\frac{\dot v}{v} C+\frac{Z}{r_{\Sigma}}\dot C
\\
-\frac{Z}{r_{\Sigma}^2}(1-2C(1+Z)+3ZC^2)=0
\end{array} \right \}
\label{II5}
\end{equation}

We shall integrate this last equation by assuming:
\begin{equation}
\left. \begin{array}{l} Z(t)=1\\
C(t)=C\equiv const
\end{array} \right \}
\label{II6}
\end{equation}
Observe that with this specific choice, this solution becomes a particular case of solution I with $C_1=Constant=0$ and $\alpha=\frac{1-C}{Cr_{\Sigma}^2}$

 Then (\ref{II5}) becomes
\begin{equation}
\ddot v+\frac{2}{r_{\Sigma}}(1-C)\dot v-\frac{1}{r_{\Sigma}^2}(1-4C+3C^2)v=0
\label{II7}
\end{equation}
whose general solution is:
\begin{equation}
v(t)=(k_1^2 e^{\beta_2 t}+k_2^2 e^{-\beta_2 t})e^{\beta_1 t}
\label{II8}
\end{equation}
with
\begin{equation}
\left. \begin{array}{l} \beta_1 =\frac{C-1}{r_{\Sigma}}\\
\beta_2 =\frac{1}{r_{\Sigma}}\sqrt{2-6C+4C^2}\\
k_1 \quad and \quad  k_2\quad  are \quad integration \quad 
constants
\end{array} \right \}
\label{II9}
\end{equation}

 Then for the velocity as defined by (\ref{19N}) we have:
\begin{equation}
U=-r\frac{[-2(1-C)\frac{\dot v}{v} -\dot C]\frac{r^2}{r_{\Sigma}^2}+[\dot C -\frac{2\dot v C}{v}]}{(1-C)\frac{r^2}{r_{\Sigma}^2}+C}
\label{velN}
\end{equation}
which, evaluated at the boundary surface gives

\begin{equation}
U_\Sigma=\frac{2r_{\Sigma}(k_1^2(\beta_2+\beta_1)
 -k_2^2(\beta_2-\beta_1) e^{-2\beta_2 t})}{k_1^2
+k_2^2 e^{-2\beta_2 t}}
\label{II10}
\end{equation}
And for the functions $C_2$ y $C_3$ we obtain:
\begin{equation}
\left. \begin{array}{l} C_3(t)=v(t)^{-2} C=\frac{C e^{-2\beta_1
t}}{(k_1^2 e^{\beta_2 t}+k_2^2 e^{-\beta_2 t})^2}\\
\\
C_2(t)=\frac{(1-C) e^{-2\beta_1 t }}{r_{\Sigma}^2 (k_1^2 e^{\beta_2
t}+k_2^2 e^{-\beta_2 t})^2}
\end{array} \right \}
\label{11II}
\end{equation}

For this solution, the physical variables become, using (\ref{IIN6}--\ref{IIN8}) and (\ref{11II}):
\begin{equation}
\mu r_{\Sigma}^2 =\frac{3}{2\pi v^4}\left[C(1-C)+r_{\Sigma}^2\frac{\dot v^2}{v^2}[(1-C)(\frac{r}{r_{\Sigma}})^2+C]^2\right]
\label{mu1}
\end{equation}

\begin{equation}
p r_{\Sigma}^2 =\frac{1-C}{2\pi v^4}\left[((1-C)(\frac{r}{r_{\Sigma}})^2-2C)+((1-C)(\frac{r}{r_{\Sigma}})^2+C)^2 (\frac{2 r_{\Sigma}\dot v}{v}+3C-1)\right]
\label{p1}
\end{equation}

\begin{equation}
q r_{\Sigma}^2 =r\frac{(1-C)}{\pi v^6}\frac{\dot v}{v}\left[[(1-C)(\frac{r}{r_{\Sigma}})^2+C]^2\right]
\label{q1}
\end{equation}

which, evaluated at the boundary surface take the form:

\begin{equation}
 \mu_{\Sigma} r_{\Sigma}^2=\frac{3}{2\pi v^{4}}\left[\frac{r_{\Sigma}^2\dot
v^2}{v^2}+C(1-C)\right]
\label{mu2}
\end{equation}
\begin{equation}
p_{\Sigma} r_{\Sigma}^2=\frac{(1-C)}{\pi v^{4}}\frac{r_{\Sigma}\dot v}{v} 
\label{p2}
\end{equation}
\begin{equation}
q_{\Sigma} r_{\Sigma}^2=\frac{(1-C)}{\pi v^{6}}\frac{r_{\Sigma}\dot v}{v} 
\label{q2}
\end{equation}

The expression for the temperature for this model may be obtained easily by putting $C_1=Constant=0$ and $\alpha=\frac{1-C}{Cr_{\Sigma}^2}$ into (\ref{2TN}), then one obtains

\begin{eqnarray}
\left. \begin{array}{l}
T(t,r)=[T_c(t)+\frac{\tau \alpha r^2}{4\pi \kappa} (\ddot C_3C_3+\dot C_3^2)
\\
\\
\qquad \qquad +\frac{\dot C_3}{4\pi
\kappa}ln(\alpha r^2+1)](\alpha r^2+1)
\end{array} \right \}
\label{2N}
\end{eqnarray}
where  $C_3(t)$ is given by (\ref{11II}). 
In this case, the system becomes a collapsing (non--dissipative) Friedmann dust sphere, when $\alpha=0$ ($C=1$). In this latter case
the temperature, as expected,  is homogeneous ($T(t)=T_c(t)$). 

The last term on the right hand side of expression (\ref{2N})  exhibits
the influence of dissipation on the  temperature,  with respect to the
non--dissipative case, as calculated from the non--causal (Landau--Eckart)
\cite{10,11}
transport equation, whereas the second term describes the contribution  of
relaxational effects. The relevance
of such effects have been brought out in recent works (see
\cite{relax} and references therein). In particular it is worth
noticing the increasing of the spatial inhomogeneity of temperature
produced by the relaxational term, an effect
which has been established before \cite{HS97}.

Let us now present a very simple model based on the solution above. The purpose here is not the modelling of any specifical astrophysical scenario, but rather to show the
feasibility of these solutions  as starting point for such modelling. 

Thus, let us consider the  following choice of constants and initial values:
\begin{equation}
k_1^2+k_2^2=1
\label{const.1}
\end{equation}
\begin{equation}
C=1+10^{-6}
\label{const.2}
\end{equation}
and
\begin{equation}
U_{\Sigma}(0)=-2.5\times 10^{-3}
\label{initial vel}
\end{equation}

Then it follows from (\ref{II9}) and (\ref{II10}) that:
\begin{equation}
\beta_1=\frac{10^{-6}}{r_{\Sigma}}
\label{beta1}
\end{equation}

\begin{equation}
\beta_2=\frac{1.4142\times 10^{-3}(1+10^{-6})}{r_{\Sigma}}
\label{beta2}
\end{equation}

\begin{equation}
k_1^2=.0577
\label{k1}
\end{equation}

\begin{equation}
k_2^2=.9423
\label{k2}
\end{equation}
With these values, it follows at once from (\ref{mu2})--(\ref{q2}) that $\mu_{\Sigma}(0)>q_{\Sigma}(0)=p_{\Sigma}(0)>0$. Furthermore since $\mu$ is a decreasing function of $r$, these
inequalities hold for all points within the sphere. As time goes on, the velocity of the boundary surface decreases (in absolute value), eventually changing of sign for a finite time
value. However physical variables remains acceptable (in this model) for values of the boundary velocity close to zero, but still negative.

Thus our model describes an initially contracting and radiating sphere, approaching the equilibrium. For later times there is a bouncing of the boundary surface, however, physical
variables become unphysical for this values of $t$, and the model  is restricted to the collapsing regime only.

\section{Conclusion}
We have presented some exact analytical  solutions to the Einstein equations, describing  spherical dissipative
shear-free and conformally flat collapse.  The  solutions are matched to the outgoing Vaidya radiating
spacetime. Besides their  simplicity, the merit of the  models resides in the fact that they
exhibit in a very clear way the influence of relaxational effects on the
temperature, and thereby on the evolution
of the system. 

In this respect we would like to stress the modifications in the temperature profile of the models, produced by the relaxational efects. This fact cannot be over emphasized. Indeed,
different temperature profiles, are not only associated with different patterns of evolution, but also, affect the luminosity profile, which is the most important element of
observational evidence in the study of dissipative collapse.

In the same line of arguments, it is worth noticing that the resulting temperature profile for each model, will depend on the specific theory of transport employed in its calculation.
Therefore, such  models might be used as test--bed for different relativistic theories of dissipation.

The specific example presented at the end of the previous section, models a dissipative collapsing configuration approaching equilibrium, with all physical variables exhibiting
appropriate behaviour. This support further our believe that the  presented solutions may be suitable for describing astrophysical scenarios involving dissipative collapsing  objects.

Finally, it is also worth noticing that  density inhomogeneities are
directly related to dissipation, while the space--time remains
conformally flat. In the non--dissipative limit ($q=L=0$), all models become homogeneous dust balls matched to Schwarzschild spacetime.

This reinforces doubts ( see for example  \cite{Bo} and references therein) on the proposal that  the Weyl tensor
\cite{Pe} or some functions of it \cite{Wa}, could  provide a gravitational
arrow of time. The rationale
behind this idea being that tidal forces tend to make the gravitating
fluid more inhomogeneous as the evolution proceeds, thereby
indicating the sense of time. However, as shown in \cite{Hetal}, density inhomogeneity, besides Weyl tensor (and the anisotropy of pressure),  also depends  on dissipation. The
solutions obtained here, clearly bring out that dependence.

\section*{Acknowledgments.}
 One of us (J.O.) gratefully acknowledges financial support from the
Spanish Ministerio de Educaci\'{o}n y Ciencia through the grant
BFM2003-02121 and from Junta de Castilla y Leon through grant SA0 10C05.


\begin{thebibliography} {99}
\bibitem{Hs} L Herrera and N O Santos  {\it Phys. Rev. D} {\bf 70}
084004 (2004).

\bibitem{Hetal} L Herrera, A  Di Prisco, J Martin, J Ospino, N O Santos
and O Troconis {\it Phys. Rev. D} {\bf 69} 084026 (2004).

\bibitem{Mitra} A Mitra {\it Phys. Rev. D} {\bf 74} 024010 (2006).

\bibitem{1} D Kazanas  and D Schramm {\it Sources of Gravitational
Radiation}, L. Smarr ed.,
(Cambridge University Press, Cambridge) (1979).

\bibitem{3} W D Arnett {\it Astrophys. J.}  {\bf 218} 815 (1977).

\bibitem{4} D Kazanas {\it Astrophys. J.} {\bf 222} L109 (1978).

\bibitem{5} J Lattimer {\it Nucl. Phys.} {\bf A478} 199 (1988).

\bibitem{8''} I Iben, {\it  Astophys. J.} 138, 1090,(1963).

\bibitem{9'} E Myra and  A.  Burrows  {\it Astrophys. J.} 364, 222, (1990).


\bibitem{18} C Cattaneo, {\it Atti. Semin. Mat. Fis. Univ. Modena},
{\bf 3}, 3 (1948).

\bibitem{12} W Israel {\it Ann. Phys., NY} {\bf 100} 310 (1976).

\bibitem{13} W Israel and J Stewart {\it Phys. Lett.}  {\bf A58} 213 (1976);
{\it Ann. Phys. NY} {\bf 118} 341 (1979).

\bibitem{15} B Carter, {\it Journ\'ees Relativistes}, Ed. M. Cahen,
Deveber R. and Geheniahau J., (ULB, Brussels) (1976).

\bibitem{14} D Pav\'on, D  Jou  and J  Casas-V\'azquez, {\it Ann. Inst. H
Poicar\'e}, {\bf A36} 79 (1982).

\bibitem{9} W  Hiscock  and L  Lindblom, {\it Ann. Phys. NY},
{\bf 151}  466 (1983).

\bibitem{7} D. Jou, J. Casas-V\'azquez J. and G. Lebon, {\it Rep. Prog. Phys.},
{\bf 51}  1105 (1988).

\bibitem{6} D  Joseph  and L  Preziosi, {\it Rev. Mod. Phys.} 
{\bf 61}  41 (1989).

\bibitem{19} J  Triginer and D  Pav\'on, {\it Class. Quantum Grav.} 
{\bf 12}  689 (1995).

\bibitem{20} D  Jou, J  Casas--Vazquez and G  Lebon, {\it Extended
Irreversible Thermodynamics}, second edition (Springer--Verlag, Berlin,
 1996).

\bibitem{21} D  Y  Tzou, {\it Macro to Micro Scale Heat Transfer:
 The Lagging Behaviour}, (Taylor \& Francis, Washington, 1996).

\bibitem{8} R  Maartens, {\it astro-ph}/9609119

\bibitem{22} A  Anile, D  Pavon and V  Romano, {\it gr--qc}/9810014. 

\bibitem{8'} L  Herrera and D  Pav\'on,
{\it Physica A}  {\bf 307}  121 (2002).

\bibitem{Bonnor} W B Bonnor, A K G de Oliveira  and N O Santos {\it
Phys. Rep.}
{\bf 181} 269 (1989).

\bibitem{relax} A Di Prisco, L Herrera and M Esculpi  {\it Class.
Quantum Grav.} {\bf 13} 1053 (1996);
A Di Prisco ,N Falc\'on, L Herrera, M Esculpi and N O Santos
{\it Gen. Rel. Grav.} {\bf 29} 1391 (1997); L Herrera and J  Mart\'\i nez 
{\it Gen. Rel. Grav.} {\bf 30} 445 (1998); M  Govender, S Maharaj and R  Maartens
{\it Class.Quantum.Grav.}
{\bf 15} 323 (1998); M Govender, R Maartens and S Maharaj {\it Month.Not.R.
Astron.Soc.} {\bf 310} 557 (1999); D Sch\"{a}fer and H F Goenner {\it Gen. Rel. Grav.}
{\bf 32} 2119 (2000);
M  Govender and K Govinder {\it Phys.Lett.A}
{\bf 283} 71 (2001); S Wagh, M Govender, K Govinder, S Maharaj, P  Muktibodh  and M Moodley 
{\it Class. Quantum Grav.} {\bf 18} 2147 (2001);
L Herrera {\it Phys. Lett. A} {\bf 300} 157 (2002); R Chan, M F A da Silva  and  J F Villas da Rocha {\it
Int. J. Modern Phys. D} {\bf 12} 347 (2003); N Naidu and M Govender {\it gr--qc/0510013};  N Naidu, M Govender and K Govinder {\it gr--qc/0509088}; L Herrera, A Di Prisco and W
Barreto {\it Phys. Rev. D} {\bf 73} 024008 (2006).


\bibitem{GM} S Maharaj and M Govender,{\it Int. J. Modern. Phys. D} {\bf 14} 667 (2005).


\bibitem{SFCF} C  B Collins and J Wainwright  {\it Phys. Rev. D}  {\bf 27} 
 1209 (1983); E  N  Glass  {\it J. Math. Phys.}  {\bf 20} 1508 (1979); P Joshi, N Dadhich and R Maartens {\it gr--qc}/0109051;
 P Joshi, R Goswami and N Dadhich {\it gr--qc}/0308012;
 L  Herrera and N O Santos {\it Month. Not. R. Astron. Soc.}, {\bf 343},
 1207 (2003).

\bibitem{Herrera}L Herrera, G Le Denmat and N O Santos {\it Int. J. Mod. Phys. D} 
{\bf 13} 583 (2004).

\bibitem{Cahill} M E Cahill  and G C McVittie {\it J. Math. Phys.}
{\bf 11} 1382 (1970).

\bibitem{MS} C W  Misner and D  H  Sharp {\it Phys. Rev.} {\bf 136}  B571 (1964).

\bibitem{Darmois}Darmois G 1927 {\it M\'{e}morial des Sciences
Math\'{e}matiques}, Gauthier-Villars, Paris, Fasc. 25

\bibitem{10}C Eckart C {\it Phys. Rev.}  {\bf 58} 919 (1940).

\bibitem{11}L Landau L and E  Lifshitz  {\it Fluid Mechanics}
(Pergamon Press, London) (1959).


\bibitem{HS97} L Herrera and N O Santos  
{\it Mon. Not. R. Astron. Soc.} {\bf 287} 161 (1997).

\bibitem{Bo} W B Bonnor 
{\it Phys. Lett.} {\bf 122A} 305 (1987);
S W Goode, A A Coley and J Wainwright 
{\it Class. Quantum Grav.} {\bf 9} 445 (1992); T Rothman {\it gr-qc/9906002}; F Mena and R Tavakol {\it Class. Quantum Grav.} {\bf 16} 435 (1999);
N Pelavas and K Lake  {\it Phys. Rev. D.} {\bf 62} 044009, (2000).

\bibitem{Pe} R Penrose  
{\it General Relativity, An Einstein Centenary Survey}
Ed. S W Hawking and W Israel (Cambridge: Cambridge University Press)
p. 581--638 (1979).

\bibitem{Wa}J Wainwright
{\it Gen. Rel. Grav.} {\bf 16} 657 (1984);
S W Goode and J Wainwright
{\it Class. Quantum Grav.} {\bf 2} 99 (1985);
W B Bonnor  
{\it Phys. Lett.} {\bf 112A} 26 (1985).
\end{thebibliography}
\end{document}